%% file: main.tex
\documentclass[%
 reprint,
 superscriptaddress,
 amsmath, amssymb,
 aps, pra,
]{revtex4-2}

\usepackage{qcircuit}
\usepackage{graphicx}
\usepackage{dcolumn}
\usepackage{bm}
\usepackage[
 colorlinks=true,
 breaklinks=true,
]{hyperref}

\begin{document}

\title{Error metric for non-trace-preserving quantum operations}
\author{Yu Shi}
 \email{shiyu@terpmail.umd.edu}
 \affiliation{Department of Electrical and Computer Engineering and Institute for Research in Electronics and Applied Physics, University of Maryland, College Park, Maryland 20742, USA}
\author{Edo Waks}
 \email{edowaks@umd.edu}
 \affiliation{Department of Electrical and Computer Engineering and Institute for Research in Electronics and Applied Physics, University of Maryland, College Park, Maryland 20742, USA}
 \affiliation{Joint Quantum Institute, University of Maryland, College Park, Maryland 20742, USA}
 \affiliation{Department of Physics, University of Maryland, College Park, Maryland 20742, USA}

\date{\today}

\begin{abstract}
We study the problem of measuring errors in non-trace-preserving quantum operations, with a focus on their impact on quantum computing. We propose an error metric that efficiently provides an upper bound on the trace distance between the normalized output states from imperfect and ideal operations, while remaining compatible with the diamond distance. As a demonstration of its application, we apply our metric in the analysis of a lossy beam splitter and a non-deterministic conditional sign-flip gate, two primary non-trace-preserving operations in the Knill-Laflamme-Milburn protocol. We then turn to the leakage errors of neutral-atom quantum computers, finding that these errors scale worse than previously anticipated, implying a more stringent fault-tolerant threshold. We also assess the quantum Zeno gate's error using our metric. In a broader context, we discuss the potential of our metric to analyze general postselected protocols, where it can be employed to study error propagation and estimate thresholds in fault-tolerant quantum computing. The results highlight the critical role of our proposed error metric in understanding and addressing challenges in practical quantum information processing.
\end{abstract}

\maketitle

\section{Introduction}

Quantifying errors in quantum operations is crucial for quantum information processing~\cite{Eisert2020, Kliesch2021}. Remarkably, fault-tolerant quantum computing necessitates the worst-case error of a quantum gate to be below a certain threshold~\cite{Aharonov2008, Knill1998, Kitaev1997}. Though numerous distance measures exist to quantify these errors~\cite{Gilchrist2005, Aharonov1998}, they primarily apply to completely positive trace-preserving linear maps. Consequently, non-trace-preserving errors, which are common in all realistic quantum systems, remain under-addressed. One prominent example of such errors is qubit leakage, observable across various quantum computing platforms, including photonics~\cite{Bongioanni2010, Shi2022, Kiesel2005, Rudolph2017}, superconducting circuits~\cite{Bultink2020, Battistel2021, Varbanov2020, Ghosh2013, Ferron2010, Chen2016, Wood2018, Motzoi2009, Herrera-Marti2013}, semiconductor spins~\cite{Andrews2019, Mehl2015, Cai2019, Byrd2005}, neutral atoms~\cite{Wu2022, Cong2022}, trapped ions~\cite{Stricker2020, Hayes2020, Brown2018, Brown2019, Brown2019a}, and topological qubits~\cite{Xu2008, Ainsworth2011}. Additionally, postselection (a technique frequently employed in quantum algorithms where qubits are manipulated based on measurement outcomes) often involves non-trace-preserving channels. These include quantum gate teleportation~\cite{Gottesman1999, Gao2010, Chou2018, Wan2019, Daiss2021}, heralded entanglement generation~\cite{Krastanov2021, Magnard2020, Hucul2015, Krutyanskiy2023, Pompili2021}, fusion gates~\cite{Bartolucci2023, Browne2005, Gimeno-Segovia2015}, quantum Zeno gates~\cite{Barontini2015, Franson2004, Raimond2012, Blumenthal2022, Kakuyanagi2015, Stannigel2014, Chen2021}, linear optics quantum gates~\cite{Knill2001, Carolan2015, Ralph2001, Okamoto2011}, measurement-based quantum computing~\cite{Raussendorf2001, Walther2005, Reimer2019, Menicucci2006}, and fault-tolerant quantum computation~\cite{Knill2004, Knill2005, Bravyi2005, Raussendorf2007, Aliferis2008}, among others. The current inability to quantify the imperfections of these channels using conventional error measures designed for trace-preserving operations underscores the need for an error metric suitable for non-trace-preserving operations.

Previous studies by Bongioanni et al.~\cite{Bongioanni2010} and Kiesel et al.~\cite{Kiesel2005} attempted to quantify the error of non-trace-preserving operations by assessing the fidelity of their normalized process matrices, a method suggested through channel-state duality~\cite{Gilchrist2005}. Nevertheless, this approach has its limitations. Notably, normalized process matrices may exhibit negative eigenvalues~\cite{Kiesel2005}, devoid of a clear physical interpretation. Additionally, this method primarily provides estimates for average-case error in quantum operations, while fault-tolerant quantum computing requires the quantification of worst-case error thresholds~\cite{Kueng2016, Sanders2016}. It's worth mentioning that the success probability serves as another metric for non-trace-preserving operations. Bongioanni et al.~\cite{Bongioanni2010} also studied this aspect, effectively characterizing it using a semidefinite operator.

Recently, Regula and Takagi~\cite{Regula2021} investigated the trade-off relations between success probability and transformation accuracy in probabilistic distillation protocols employing postselection. Concurrently, Gavorová~\cite{Gavorova2020} introduced a modified version of the diamond distance~\cite{Aharonov1998} for postselected computations, additionally providing a metric to bound it. However, the metric is constrained by two major limitations. First, it can only calculate the distance between non-trace-preserving and trace-preserving operations. Second, it fails to converge to the conventional diamond distance when the operation is infinitesimally non-trace-preserving, thereby offering only a loose bound on the distance.

In this paper, we propose a general metric designed to compare two non-trace-preserving quantum operations. This metric offers an efficient means to compute a worst-case error bound and addresses the previously mentioned limitations. We demonstrate its utility by applying it to the analysis of a lossy beam splitter and a non-deterministic conditional sign-flip gate in the Knill-Laflamme-Milburn (KLM) protocol~\cite{Knill2001}. Furthermore, we use our metric to examine the leakage errors in neutral-atom quantum computers, revealing a much worse error scaling than previously estimated by the Pauli twirling approximation~\cite{Wu2022}. Such insight suggests a tighter threshold for fault-tolerant quantum computing. Additionally, we apply our metric to a quantum Zeno gate on transmons~\cite{Blumenthal2022} to quantify the error in the presence of non-ideal measurements. These applications underscore the versatility and superiority of our metric. We further discuss the potential of our metric for use in general postselected protocols, focusing on its utility for studying error propagation and estimating thresholds in fault-tolerant quantum computing. Collectively, our findings not only broaden the scope of quantifiable imperfections in quantum computers but also aid in designing more reliable and efficient quantum error-correcting schemes.

\section{Review}

\subsection{Diamond distance}

We begin by revisiting the essential properties a measure should possess to effectively distinguish between different quantum operations. A quantum operation $\mathcal{E}$ acts as a map between input and output states, represented by $\rho_{\rm out}=\mathcal{E}\left(\rho_{\rm in}\right)$. The distance, denoted as $\Delta\left(\mathcal{E},\mathcal{F}\right)$ between quantum operations $\mathcal{E}$ and $\mathcal{F}$, quantifies the differences in their output states given identical input states. Any distance measure should adhere to certain mathematical properties as outlined by Gilchrist et al.~\cite{Gilchrist2005}.
\begin{enumerate}
    \item \textit{Metric}. This property necessitates the fulfillment of three conditions: (a) non-negativity, $\Delta\left(\mathcal{E},\mathcal{F}\right)\ge0$, becoming zero exclusively when $\mathcal{E}=\mathcal{F}$; (b) symmetry, $\Delta\left(\mathcal{E},\mathcal{F}\right)=\Delta\left(\mathcal{F},\mathcal{E}\right)$; and (c) triangle inequality, $\Delta\left(\mathcal{E},\mathcal{F}\right)\le\Delta\left(\mathcal{E},\mathcal{G}\right)+\Delta\left(\mathcal{G},\mathcal{F}\right)$.
    \item \textit{Chaining}. Represented as $\Delta\left(\mathcal{E}_2\circ\mathcal{E}_1,\mathcal{F}_2\circ\mathcal{F}_1\right)\le\Delta\left(\mathcal{E}_1,\mathcal{F}_1\right)+\Delta\left(\mathcal{E}_2,\mathcal{F}_2\right)$. Here, $\mathcal{E}_2\circ\mathcal{E}_1$ indicates a composite process where operation $\mathcal{E}_1$ precedes operation $\mathcal{E}_2$ (analogously for $\mathcal{F}$). This property ensures the cumulative distance for combined operations remains less than the sum of their individual distances.
    \item \textit{Stability}. Defined by $\Delta\left(\mathcal{E}\otimes\mathcal{I},\mathcal{F}\otimes\mathcal{I}\right)=\Delta\left(\mathcal{E},\mathcal{F}\right)$, where \( \mathcal{I} \) denotes the identity operation in an additional state space. This property guarantees that the distance remains invariant when the operations act on a subsystem, even if this subsystem is entangled with external systems.
\end{enumerate}
In practical contexts, our primary interest often centers on the deviation between an experimentally realized quantum operation and its ideal counterpart, with the distance measure serving as an error metric. Both the chaining and stability properties are invaluable when assessing extensive quantum information tasks composed of sequentially applied quantum operations. By leveraging these properties, we can estimate the error of a comprehensive process based on the cumulative distances of its sequential individual operations.

The diamond distance~\cite{Aharonov1998} is a widely used measure to distinguish between two quantum channels, i.e., completely positive trace-preserving linear maps. It quantifies the maximum trace distance between the output states of two quantum channels given the same input states. The diamond distance $d_\diamond\left(\mathcal{E},\mathcal{F}\right)$ between quantum channels $\mathcal{E}$ and $\mathcal{F}$ is defined as
\begin{equation} \label{eq:diamond_distance}
d_\diamond\left(\mathcal{E},\mathcal{F}\right)=\frac{1}{2}\max_\rho{\left\|\left(\mathcal{E}\otimes\mathcal{I}_R\right)\left(\rho\right)-\left(\mathcal{F}\otimes\mathcal{I}_R\right)\left(\rho\right)\right\|_1}\;.
\end{equation}
In the above equation, $\left\|\cdot\right\|_1$ denotes the trace norm. The quantum channels $\mathcal{E}$ and $\mathcal{F}$ map quantum states from input space $X$ to output space $Y$. $\mathcal{I}_R$ represents the identity operation on an auxiliary space $R$, and $\rho$ stands for a density operator on the composite space $X\otimes R$. Significantly, the diamond distance satisfies all three requisite properties previously detailed for an effective distance measure. Its metric property arises directly from the trace norm. The chaining property can be derived from the triangle inequality of the trace norm. The stability property holds under the condition $dim\left(X\right)\le dim\left(R\right)$~\cite{Watrous2018}. Although deriving an analytical expression for the diamond distance of general quantum channels is challenging, it can be computed using convex optimization~\cite{Watrous2009}.

A natural extension of the diamond distance for non-trace-preserving operations is to measure the maximum trace distance between their normalized output states. This extended distance, referred to as the “general distance” and denoted as $d_g$, is defined by
\begin{widetext}
\begin{equation} \label{eq:general_distance}
d_g\left(\mathcal{E},\mathcal{F}\right)=\frac{1}{2}\max_\rho{\left\|\frac{\left(\mathcal{E}\otimes\mathcal{I}_R\right)\left(\rho\right)}{Tr\left[\left(\mathcal{E}\otimes\mathcal{I}_R\right)\left(\rho\right)\right]}-\frac{\left(\mathcal{F}\otimes\mathcal{I}_R\right)\left(\rho\right)}{Tr\left[\left(\mathcal{F}\otimes\mathcal{I}_R\right)\left(\rho\right)\right]}\right\|_1}\;.
\end{equation}
\end{widetext}
Parameters here have the same definitions as in Equation~\ref{eq:diamond_distance}, with the distinction that quantum operations $\mathcal{E}$ and $\mathcal{F}$ can be non-trace-preserving. This means that $Tr\left[\mathcal{E}\left(\rho\right)\right] \le 1$ and $Tr\left[\mathcal{F}\left(\rho\right)\right] \le 1$. When quantum operations $\mathcal{E}$ and $\mathcal{F}$ are trace-preserving, the equalities hold, and the general distance reduces to the conventional diamond distance. However, applying the general distance in practice presents a challenge. Unlike the diamond distance, which can be computed efficiently, evaluating Equation~\ref{eq:general_distance} is not straightforward since it does not represent a convex problem. Consequently, there is a need for a metric that provides an upper bound for the general distance.

\subsection{Gavorová's metric}

Gavorová~\cite{Gavorova2020} studied Equation~\ref{eq:general_distance} in the context of postselected computations and proposed an upper bound that serves as a metric:
\begin{equation} \label{eq:Gavorova}
d_g\left(\mathcal{E},\mathcal{F}\right)\le2d_\diamond\left(\frac{\mathcal{E}}{\left\|\mathcal{E}\right\|_1},\mathcal{F}\right)\;,
\end{equation}
where $\left\|\mathcal{E}\right\|_1$ is the induced trace norm~\cite{Watrous2018} of operation $\mathcal{E}$. The term $d_\diamond\left(\frac{\mathcal{E}}{\left\|\mathcal{E}\right\|_1},\mathcal{F}\right)$ represents the diamond distance between operations without normalizing their output states. This metric is computable using convex optimization. Nonetheless, it has two main limitations. First, Equation~\ref{eq:Gavorova} requires operation $\mathcal{F}$ to be trace-preserving. As a result, it cannot be applied to compute the general distance between two non-trace-preserving operations. Second, as operations $\mathcal{E}$ and $\mathcal{F}$ approach the asymptotic limit of being trace-preserving, an ideal metric should converge to the diamond distance $d_\diamond\left(\mathcal{E},\mathcal{F}\right)$. However, Equation~\ref{eq:Gavorova} does not meet this expectation, providing a rather loose bound on the general distance even when an operation is only infinitesimally non-trace-preserving.

\section{Main Result}

Inspired by renormalization techniques~\cite{White1992, Verstraete2004, Vidal2007, Evenbly2015}, we develop a metric that provides an upper bound on the general distance, as defined in Equation~\ref{eq:general_distance}. Let us consider two quantum operations expressed in the operator-sum representation:
\begin{eqnarray*}
\mathcal{E}\left(\rho\right)=\sum_{i}{A_i\rho A_i^\dag}\;,\\
\mathcal{F}\left(\rho\right)=\sum_{j}{B_j\rho B_j^\dag}\;,
\end{eqnarray*}
where $A_i$ and $B_j$ are Kraus operators that map the quantum state vector from input space $X$ to output space $Y$. These operators adhere to the conditions $\sum_{i}{A_i^\dag A_i}\le I$ and $\sum_{j}{B_j^\dag B_j}\le I$. We define the dimension of $X$ as $m$, the dimension of $Y$ as $n$, and the cardinalities of sets $\{A_i\}$ and $\{B_j\}$ as $k$ and $k^\prime$, respectively.

The derivation of our metric begins by representing these quantum operations using the Stinespring dilation~\cite{Stinespring1955}:
\begin{eqnarray*}
\mathcal{E}\left(\rho\right)=Tr_Z\left\{\hat{A}\rho\hat{A}^\dag\right\}\;,\\
\mathcal{F}\left(\rho\right)=Tr_Z\left\{\hat{B}\rho\hat{B}^\dag\right\}\;,
\end{eqnarray*}
Here, vectors $e_i$ and $e_j$ denote the bases on the environmental space $Z$. The isometries $\hat{A}$ and $\hat{B}$ are defined by $\hat{A}=\sum_{i}{e_i\otimes A_i}$ and $\hat{B}=\sum_{j}{e_j\otimes B_j}$, respectively. Given that Equation~\ref{eq:general_distance} maximizes over a quantum state on the composite space $X\otimes R$, it is necessary to extend both isometries to $\hat{A}\otimes I_R$ and $\hat{B}\otimes I_R$. This extension ensures that the isometries capture the dynamics of the entire system, mapping quantum states from input space $X\otimes R$ to output space $Z\otimes Y\otimes R$. As detailed in Appendix~\ref{app:stability}, we demonstrate that Equation~\ref{eq:general_distance} satisfies the stability property, such that $d_g\left(\mathcal{E}\otimes\mathcal{I},\mathcal{F}\otimes\mathcal{I}\right)=d_g\left(\mathcal{E},\mathcal{F}\right)$ if $dim\left(X\right)\le dim\left(R\right)$. Under this condition, the optimization variable can be simplified from a density operator to a pure state. Consequently, we can reformulate Equation~\ref{eq:general_distance} as:
\begin{equation} \label{eq:isometry}
d_g\left(\mathcal{E},\mathcal{F}\right)=\frac{1}{2}\max_\psi{\left\|Tr_Z\left\{\frac{\hat{A}\psi\psi^\dag\hat{A}^\dag}{\psi^\dag\hat{A}^\dag\hat{A}\psi}-\frac{\hat{B}\psi\psi^\dag\hat{B}^\dag}{\psi^\dag\hat{B}^\dag\hat{B}\psi}\right\}\right\|_1}\;,
\end{equation}
where $\psi$ is a pure state on the extended input space $X\otimes R$. For conciseness, we use $\hat{A}$ as a shorthand for $\hat{A}\otimes I_R$ and $\hat{B}$ for $\hat{B}\otimes I_R$.

To bound Equation~\ref{eq:isometry}, we employ the singular value decomposition (SVD). By applying SVD to $\hat{B}\hat{A}^{-1}$, we obtain $\hat{B}\hat{A}^{-1}=UMV^\dag$. Here, $U$ and $V$ are isometries mapping from input space $X$ to output space $Z\otimes Y$. Specifically, $U$ is of dimension $k^\prime n\times m$, $V$ is $kn\times m$, and $M$ is a nonnegative diagonal operator with a dimension of $m\times m$. Building on these isometries, we introduce two renormalized trace-preserving operations
\begin{eqnarray}
    \mathcal{U}\left(\rho\right)&=&Tr_Z\left\{U\rho U^\dag\right\}\;,\nonumber\\
    \mathcal{V}\left(\rho\right)&=&Tr_Z\left\{V\rho V^\dag\right\}\;,
\end{eqnarray}
and a normalization operation defined as
\begin{equation}
    \mathcal{M}\left(\rho\right)=\frac{M\rho M}{Tr\left[M\rho M\right]}\;.
    \label{eq:normalization}
\end{equation}
With these definitions in place, we illustrate in Appendix~\ref{app:renormalization} that Equation~\ref{eq:isometry} can be bounded by
\begin{equation}
    d_g\left(\mathcal{E},\mathcal{F}\right)\le d_\diamond\left(\mathcal{U},\mathcal{V}\right)+d_g\left(\mathcal{I},\mathcal{M}\right)\;.
    \label{eq:bound}
\end{equation}
Here, $d_\diamond\left(\mathcal{U},\mathcal{V}\right)$ denotes the conventional diamond distance, which can be computed using convex optimization. On the other hand, $d_g\left(\mathcal{I},\mathcal{M}\right)$ represents the general distance between an identity operation and the normalization operation, termed the normalization distance. Analytically, this distance is given by
\begin{equation}
    d_g\left(\mathcal{I},\mathcal{M}\right)=\frac{\lambda_{\rm max}-\lambda_{\rm min}}{\lambda_{\rm max}+\lambda_{\rm min}}\;,
    \label{eq:normalization_distance}
\end{equation}
where $\lambda_{\rm max}$ and $\lambda_{\rm min}$ are the maximum and minimum eigenvalues of $M$, respectively. (See Appendix~\ref{app:normalization_distance} for a detailed derivation.) Thus, Equation~\ref{eq:bound} provides an efficiently computable upper bound for the general distance, serving as a metric for quantifying errors in non-trace-preserving operations.

While our bound might not always be tight in all cases, we advocate its efficacy as an error metric for non-trace-preserving quantum operations for several reasons: (a) It provides an upper bound on the worst-case error, aiding in determining thresholds for fault-tolerant quantum computing and evaluating quantum system performance. (b) The bound can be efficiently computed and applied to compare any two non-trace-preserving quantum operations. (c) The discrepancy between our metric and the exact distance remains within the same order of magnitude. Moreover, this gap diminishes as physical imperfections vanish. (d) The gap arises from applying the triangle inequality to the sum of the distances for two sequential operations, which is consistent with the chaining property expected of a robust distance measure. In the context of large quantum systems, this suggests that the total distance can be approximated by the sum of individual steps. (e) If the operations under comparison exhibit only infinitesimal deviations from being trace-preserving, the gap vanishes. Given these advantages, we are confident in the potential of our metric for analyzing quantum operations.

\section{Applications}

In this section, we explore several concrete examples to clarify the utility of our metric. First, we analyze the two primary non-trace-preserving mechanisms for the Knill-Laflamme-Milburn (KLM) protocol~\cite{Knill2001}: the unbalanced lossy beam splitter, used for single-qubit rotation, and the non-deterministic conditional sign-flip (NS) gate, operating in the presence of detector dark counts. When combined with trace-preserving mechanisms like phase instability, these results provide a complete error analysis for linear optical quantum computing. Second, we apply our metric to examine the detect-and-reset method intended for leakage reduction in neutral-atom quantum computers~\cite{Wu2022}. Contrary to previous beliefs, our findings indicate that this approach fails to fully mitigate the leakage effect, thus necessitating a stricter fault-tolerant threshold. Third, we direct our attention to a recent quantum Zeno gate proposed by Blumenthal et al.~\cite{Blumenthal2022} for transmons. Our metric presents an exact error measure for the imperfect quantum Zero gate, substantially outperforming the ad-hoc metric in the original paper. Additionally, we discuss the application of our metric in other postselected protocols, such as quantum teleportation and cluster state quantum computing, with details further elaborated in Appendix~\ref{app:examples}. For our analyses, we employ numerical simulations using the PICOS~\cite{PICOS2022} convex optimization package in Python.

\subsection{Lossy beam splitter}

We begin by examining the single-qubit rotation, which is implemented using a beam splitter. Faulty beam splitters, characterized by state-dependent photon loss, can reduce the success probability of the KLM protocol and also induce incorrect rotations to the quantum state. The operator associated with a beam splitter, denoted as $B_{\theta,\phi}$, is expressed by
\begin{equation}
U\left(B_{\theta,\phi}\right)=\begin{pmatrix}\gamma_t\cos{\theta}&-\gamma_r\sin{\theta}e^{i\phi}\\\gamma_r\sin{\theta}e^{-i\phi}&\gamma_t\cos{\theta}\end{pmatrix}\;,
\end{equation}
where $\gamma_r$ and $\gamma_t$ represent the photon loss rates for reflection and transmission, respectively. If $\gamma_r,\gamma_t=1$, the operator corresponds to an ideal beam splitter, denoted as $U_{\rm id}$. Conversely, we denote the operator associated with a faulty beam splitter as $U_r$. The distance between the ideal and faulty beam splitters, after postselecting the non-loss instances, is
\begin{equation}
d_g\left(U_{\rm id},U_r\right)=\frac{1}{2}\max_\rho{\left\|U_{\rm id}\rho U_{\rm id}^\dag-\frac{U_r\rho U_r^\dag}{Tr\left[U_r\rho U_r^\dag\right]}\right\|_1}\;,
\end{equation}
where we abbreviate $U_k\otimes I$ as $U_k$ for conciseness. Using Equation~\ref{eq:bound}, we compute this distance to be
\begin{equation} \label{eq:beamsplitter}
d_g\left(U_{\rm id},U_r\right)=\frac{\left|\gamma_r-\gamma_t\right|\cos{\theta}\sin{\theta}}{\sqrt{\gamma_t^2\cos^2{\theta}+\gamma_r^2\sin^2{\theta}}}\;.
\end{equation}
Note that the above metric is exact, rather than an upper bound. This is because the normalization operator \( M=\sqrt{\frac{\gamma_r^2+\gamma_t^2}{2}}I \) leads to a normalization distance of zero, as indicated by Equation~\ref{eq:normalization_distance}.

\begin{figure}[tb]
    \centering
    \includegraphics[width=1.0\columnwidth]{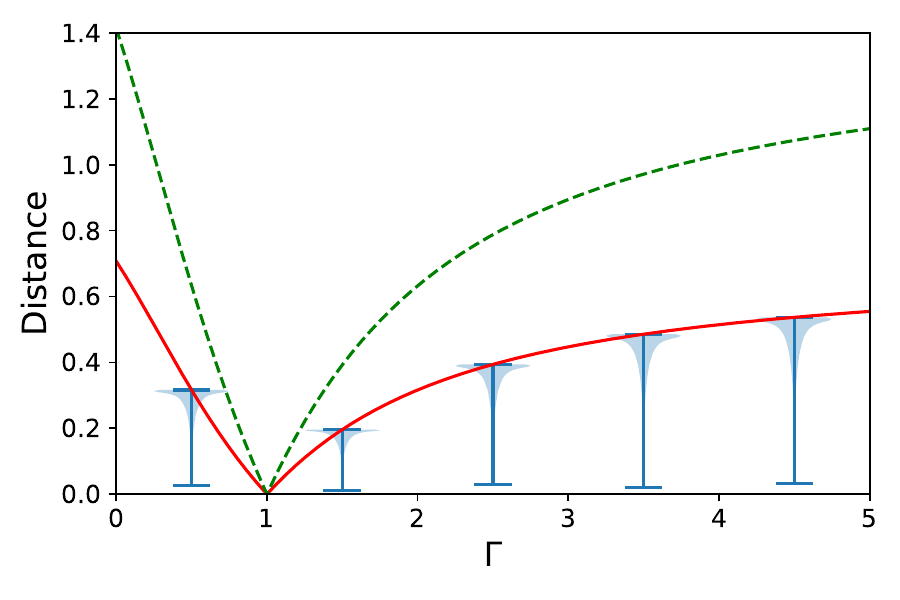}
    \caption{Distance between ideal and faulty beam splitters as a function of photon loss ratio $\Gamma=\gamma_r/\gamma_t$. The solid red line represents our metric, the dashed green line indicates Gavorová's bound, and the blue violin plots depict trace distance statistics from the Monte Carlo simulation.}
    \label{fig:bs}
\end{figure}

Figure~\ref{fig:bs} depicts the general distance between the ideal and faulty beam splitters, with the solid red line representing the distance calculated by Equation~\ref{eq:beamsplitter} for $\theta=\pi/4$. This condition corresponds to a $50/50$ beam splitter in the ideal scenario. The general distance is plotted as a function of $\Gamma=\gamma_r/\gamma_t$, signifying the ratio of photon loss between reflection and transmission. As anticipated, the distance attains an ideal value of zero when the loss is balanced ($\gamma_r=\gamma_t$). To ascertain that the calculated distance is exact, we compare it with a Monte Carlo simulation. In this simulation, we apply operator $U\left(B_{\pi/4,0}\right)$ to randomly generated input states and compute the trace distance between the normalized output states from both the ideal and faulty beam splitters. The blue violin plots in Figure~\ref{fig:bs} illustrate the statistics of these trace distances. The vertical segments of the plots delineate the range of trace distances, while their width reveals the probability distribution. Most of the simulated trace distances are concentrated near their maximum, corresponding to the value calculated by Equation~\ref{eq:beamsplitter}. This alignment indicates that the theoretical calculation is accurate and robust. Furthermore, we make a comparative assessment of our metric against the one proposed by Gavorová, as computed by Equation~\ref{eq:Gavorova}. This metric is plotted as the dashed green line in Figure~\ref{fig:bs}, demonstrating that it approximates twice the actual general distance.

\subsection{Non-deterministic conditional sign-flip gate}

We next analyze the non-deterministic conditional sign-flip (NS) gate, a quantum gate that applies a non-linear phase shift to a single photonic mode by utilizing two ancilla modes. We investigate a system where the initial state is prepared in the form $\left|\psi_{\rm in}\right\rangle\otimes\left|10\right\rangle_A$, where $\left|\psi_{\rm in}\right\rangle=\left(\ \alpha_0\left|0\right\rangle+\alpha_1\left|1\right\rangle+\alpha_2\left|2\right\rangle\right)$ represents the input state in the photon number basis, and $\left|10\right\rangle_A$ signifies the state of the two ancilla modes. The NS gate, upon applying linear optical transformations and detecting the ancilla modes in state $\left|10\right\rangle_A$, yields the output state $\left|\psi_{\rm out}\right\rangle=\left(\alpha_0\left|0\right\rangle+\alpha_1\left|1\right\rangle-\alpha_2\left|2\right\rangle\right)$. However, this gate is susceptible to errors due to detector dark counts, which can result in incorrect detection events.

To employ our metric in this specific context, we define the state space for $\left|\psi_{\rm in}\right\rangle$ as $X=\{\left|0\right\rangle,\left|1\right\rangle,\left|2\right\rangle\}$ and for $\left|\psi_{\rm out}\right\rangle$ as $Y=\{\left|0\right\rangle,\left|1\right\rangle,\left|2\right\rangle,\left|3\right\rangle\}$. The inclusion of an additional photon in the output space is necessary, as the ancilla photon might mistakenly exit into the output port due to dark counts triggering a detection event in the first ancilla mode. The ideal NS gate transforms the input state as 
\begin{equation}
\mathcal{NS}_{\rm id}\left(\rho\right)=\frac{E_{10}\rho E_{10}^\dag}{\text{Tr}\left[E_{10}\rho E_{10}^\dag\right]}\;,
\end{equation}
and the faulty NS gate transforms the state as
\begin{equation}
\mathcal{NS}_r\left(\rho\right)=\frac{E_{10}\rho E_{10}^\dag+\mu E_{00}\rho E_{00}^\dag}{\text{Tr}\left[E_{10}\rho E_{10}^\dag+\mu E_{00}\rho E_{00}^\dag\right]}\;,
\end{equation}
where $\mu$ represents the dark count rate. The operators $E_{10}$ and $E_{00}$, conditioned on the measurement outcomes of the ancilla modes $\left|10\right\rangle_A$ and $\left|00\right\rangle_A$ respectively~\cite{Knill2001}, are given by
\begin{eqnarray*}
    E_{10}&=&\begin{pmatrix}\frac{1}{2}&0&0\\0&\frac{1}{2}&0\\0&0&-\frac{1}{2}\\0&0&0\end{pmatrix}\;,\\
    E_{00}&=&\begin{pmatrix}0&0&0\\\frac{1}{2^{1/4}}&0&0\\0&\frac{-2+\sqrt2}{2^{1/4}}&0\\0&0&\sqrt6\left(\frac{3}{2^{3/4}}-2^{3/4}\right)\end{pmatrix}\;.
\end{eqnarray*}
Within these parameters, the general distance between an ideal NS gate and a faulty one, in the presence of detector dark counts, can be expressed as
\begin{equation}
    d_g\left(\mathcal{NS}_{\rm id},\mathcal{NS}_r\right)=\frac{1}{2}\max_\rho{\left\|\mathcal{NS}_{\rm id}\left(\rho\right)-\mathcal{NS}_r\left(\rho\right)\right\|_1}\;,
\end{equation}
where $\mathcal{NS}_k\otimes\mathcal{I}$ is abbreviated as $\mathcal{NS}_k$ for conciseness.

\begin{figure}[tb]
    \centering
    \includegraphics[width=1.0\columnwidth]{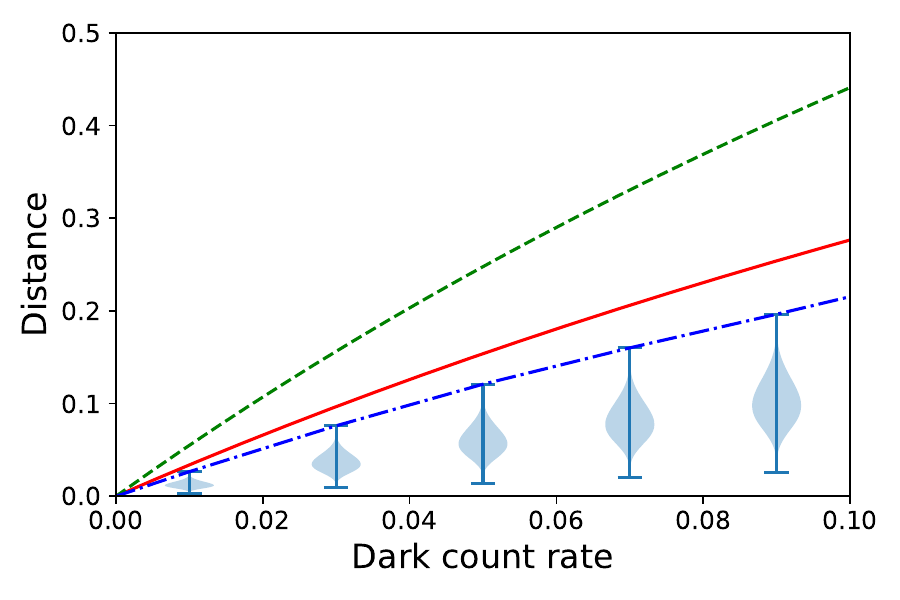}
    \caption{Distance between ideal and faulty non-deterministic conditional sign-flip (NS) gates as a function of dark count rate. The solid red line represents our metric, the dashed green line indicates Gavorová's bound, the blue violin plots depict trace distance statistics from the Monte Carlo simulation, and the dash-dotted blue line highlights the maximal trace distance obtained from the simulation.}
    \label{fig:ns}
\end{figure}

Figure~\ref{fig:ns} depicts the distance as a function of the dark count rate. The solid red line represents our metric for the quantum operation distance between the ideal and faulty NS gates, as calculated by Equation~\ref{eq:bound}. The blue violin plots illustrate the statistics of the trace distance between the output states from a Monte Carlo simulation. In this simulation, we apply the ideal and faulty NS gate to randomly generated input states and calculate the trace distance between the normalized output states. Most of the probability is concentrated around the median of the distribution. The dash-dotted blue line signifies the maximal trace distance from the Monte Carlo simulation, reflecting the actual general distance. While Equation~\ref{eq:bound} indeed provides an upper bound on the distance in this case, the bound is no longer tight. The lack of tightness can be discerned from the gap between the dash-dotted and solid lines. Finally, the dashed green line illustrates the metric by Gavorová, also standing as an upper bound on the actual error but being even looser than our result.

\subsection{Leakage in neutral atoms}

In our third example, we turn our attention to the analysis of leakage errors in neutral-atom quantum computers. Leakages occur when qubits leave the computational subspace. In neutral-atom quantum computers, qubits are encoded in metastable states. To entangle qubits, the atoms are typically driven to highly excited Rydberg states, which can spontaneously decay to irrelevant low-energy states and lead to leakages. Such leakages constitute the dominant errors of neutral atoms~\cite{Cong2022, Wu2022} and are highly problematic. They corrupt the subspace encoding and invalidate the standard quantum error-correcting codes with syndrome measurements, which only provide information about error events restricted to a subspace~\cite{Byrd2005}.

One strategy to address leakage errors involves employing the detect-and-reset method~\cite{Knill1996, Preskill1997, Bultink2020}. Within this approach, each qubit undergoes repeatedly monitoring for leakages. If a leak event is detected, the affected qubit is discarded, replaced with an initialized state in the computational subspace, and subsequently recovered using a standard error-correcting scheme. This method was examined by Wu et al.~\cite{Wu2022} in the context of neutral-atom quantum computers. They assumed that each atom was subject to a leakage probability of $p$. When a leak event was detected, they reset the qubit to state $\left|0\right\rangle$ and recover it by the XZZX surface code~\cite{Bonilla2022}. However, if a no-leak event was detected, the qubit was subjected to a non-trace-preserving operation defined as
\begin{equation}
\mathcal{E}_K\left(\rho\right)=\frac{K\rho K^\dag}{Tr\left[K\rho K^\dag\right]}\;,
\end{equation}
where $K=I+\left(\sqrt{1-p}-1\right)\left|1\middle\rangle\middle\langle1\right|$ represented the no-leak operator. By applying the Pauli twirling approximation~\cite{Silva2008} to operator $K$, the authors approximated the operation as a channel imposing a Pauli-Z error at a rate proportional to $p^2$. The error is then assumed to be negligible because it scales quadratically, while other leakage errors due to detection inefficiencies scale linearly. With this approximation, Wu et al. concluded that $98\%$ of leakage errors could be corrected. However, the Pauli twirling approximation is typically applicable only to trace-preserving linear maps (as detailed in Appendix~\ref{app:PTA}), highlighting the need to reexamine this analysis using a metric compatible with non-trace-preserving operations.

We proceed to apply our metric to this model. The undetected leakage channel can be expressed as
\begin{equation}
\mathcal{E}_{\rm leak}\left(\rho\right)=K\rho K^\dag+L\rho L^\dag\;,
\end{equation}
where $K$ represents the no-leak operator, and $L=\sqrt{p}\left|e\middle\rangle\middle\langle1\right|$ represents the operator associated with the leakage of a qubit from computational state $\left|1\right\rangle$ to an irrelevant state $\left|e\right\rangle$. The error of this undetected leakage channel can be quantified by the diamond distance $d_{\rm leak}=d_\diamond\left(\mathcal{E}_{\rm leak},\mathcal{I}\right)$, where $\mathcal{I}$ denotes the identity channel, resulting in $d_{\rm leak}=p$. When applying the detect-and-reset method with a detection efficiency of $\gamma$, the postselected channel is described as
\begin{equation}
\mathcal{E}_{\rm post}\left(\rho\right)=\frac{K\rho K^\dag+\left(1-\gamma\right)L\rho L^\dag}{Tr\left[K\rho K^\dag+\left(1-\gamma\right)L\rho L^\dag\right]}\;,
\end{equation}
The corresponding error should be evaluated by the general distance $d_{\rm post}=d_g\left(\mathcal{E}_{\rm post},\mathcal{I}\right)$. In the limit of perfect detection efficiency ($\gamma=1$), $\mathcal{E}_{\rm post}$ aligns with the no-leak channel solely described by operator $K$. We can calculate the distance by Equation~\ref{eq:normalization}, yielding 
\begin{equation}
d_{\rm post}=\frac{1-\sqrt{1-p}}{1+\sqrt{1-p}}\;.
\end{equation}
Notably, this distance is proportional to $p$ rather than $p^2$. Therefore, it cannot be regarded as negligible and must be included in the error analysis.

\begin{figure}[tb]
    \centering
    \includegraphics[width=1.0\columnwidth]{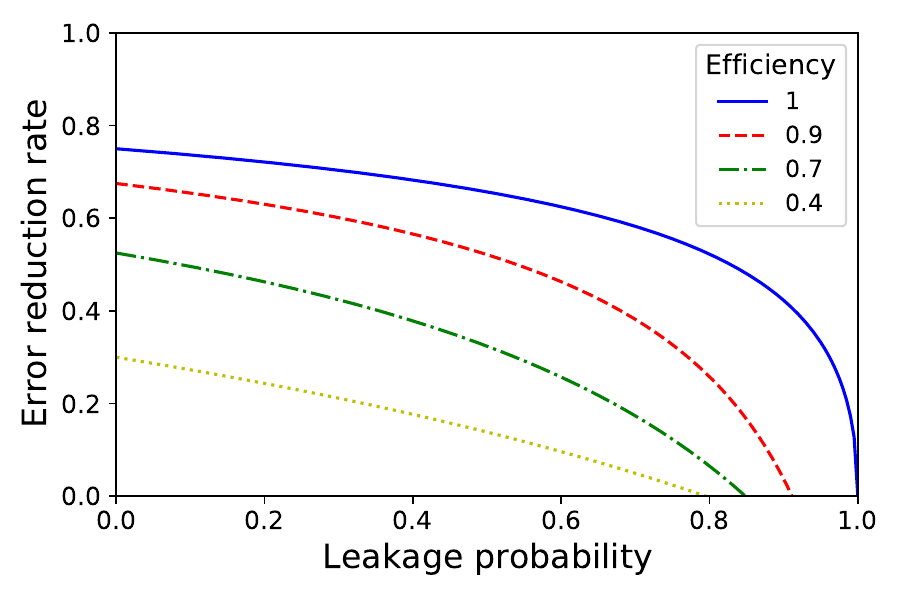}
    \caption{Error reduction rate as a function of leakage probability for different detection efficiencies.}
    \label{fig:conversion}
\end{figure}

For a comprehensive analysis that accounts for both detection inefficiencies and errors due to the no-leak evolution, we define the error reduction rate $\eta=1-\frac{d_{\rm post}}{d_{\rm leak}}$. Figure~\ref{fig:conversion} illustrates this rate as a function of leakage probability across various detection efficiencies. These plots establish a lower bound on the error reduction for the given detection efficiency (refer to Appendix~\ref{app:leakge_reduction} for detailed calculations). Specifically, the calculated error reduction rate with ideal detection efficiency is exact. Therefore, it imposes a stringent upper limit on all potential error reductions attributable to the detect-and-reset method, resulting in an optimal rate of $75\%$. This finding contrasts to the $98\%$ reduction predicted by the authors of~\cite{Wu2022}, which was presumed to be constrained solely by detection inefficiencies.

\subsection{Imperfect quantum Zeno gates}

In the final example, we explore the application of our metric to quantum Zeno gates, a prominent example of postselected protocols. The quantum Zeno effect~\cite{Facchi2008} confines a quantum system within selected subspaces through measurement, thereby generating non-trivial multi-qubit dynamics. Ideally, a quantum Zeno gate would achieve a success probability of $1$, assuming continuous measurements. However, measurements are not always present in practice, rendering the gate probabilistic. These non-ideal conditions can cause non-trace-preserving errors, emphasizing the need for a comprehensive analysis.

As a concrete example, we analyze a protocol recently proposed by Blumenthal et al.~\cite{Blumenthal2022}, which demonstrates an entangling gate on transmons (superconducting qubits) using the quantum Zeno effect. They consider two transmons with eigenstates $\left|g\right\rangle$ and $\left|e\right\rangle$, representing the computational bases. Additionally, the first transmon possesses an extra eigenstate $\left|f\right\rangle$ as an ancillary state. The entangling gate between the two transmons is implemented by driving the first transmon at a Rabi frequency $\Omega_R$ between states $\left|e\right\rangle$ and $\left|f\right\rangle$, with the Hamiltonian expressed as
\begin{equation}
    \mathcal{H}=\frac{1}{2}i\Omega_R\left(\left|e\middle\rangle\middle\langle f\right|-\left|f\middle\rangle\middle\langle e\right|\right)\otimes I\;,
\end{equation}
where $I$ represents the identity operator on the second transmon. Meanwhile, rapid measurements are applied to the two transmons using the projector $P_0=I-\left|fe\middle\rangle\middle\langle fe\right|$. In the limit of infinitely rapid projections, the Zeno effect dressed Hamiltonian $P_0\mathcal{H}P_0$ is expressed as
\begin{eqnarray}
    \mathcal{H}_Z = i\frac{\Omega_R}{2}\left(\left|eg\middle\rangle\middle\langle fg\right|-\left|fg\middle\rangle\middle\langle ef\right|\right)\;,
\end{eqnarray}
resulting in an unitary of $U_Z\left(t\right)=P_0e^{-it\mathcal{H}_Z}P_0$. At $t=\frac{2\pi}{\Omega_R}$, the unitary is a controlled-phase gate given by
\begin{equation}
    U_Z=\left|gg\middle\rangle\middle\langle gg\right|+\left|ge\middle\rangle\middle\langle ge\right|-\left|eg\middle\rangle\middle\langle eg\right|+\left|ee\middle\rangle\middle\langle ee\right|\;.
\end{equation}
With a finite measurement rate of $\Gamma$, the Zeno gate is no longer unitary and can be modeled as
\begin{equation} \label{eq:Zeno_transformation}
    \mathcal{E}_Z\left(\rho\right)=P_0e^{t\left(\Gamma\mathcal{D}-i\mathcal{H}\right)}P_0\rho P_0e^{t\left(\Gamma\mathcal{D}+i\mathcal{H}\right)}P_0\;,
\end{equation}
where $\mathcal{D}=-\frac{1}{2}\left|fe\middle\rangle\middle\langle fe\right|$ represents a dissipation operator (refer to Supplementary Section 1 in~\cite{Blumenthal2022} for details). This imperfect Zeno gate exhibits a non-trace-preserving error that can be quantified by
\begin{equation} \label{eq:Zeno}
d_g\left(\mathcal{E}_Z,U_Z\right)=\frac{1}{2}\max_\rho{\left\|\frac{\mathcal{E}_Z\left(\rho\right)}{Tr\left[\mathcal{E}_Z\left(\rho\right)\right]}-U_Z\rho U_Z^\dag\right\|_1}\;.
\end{equation}

\begin{figure}[tb]
    \centering
    \includegraphics[width=1.0\columnwidth]{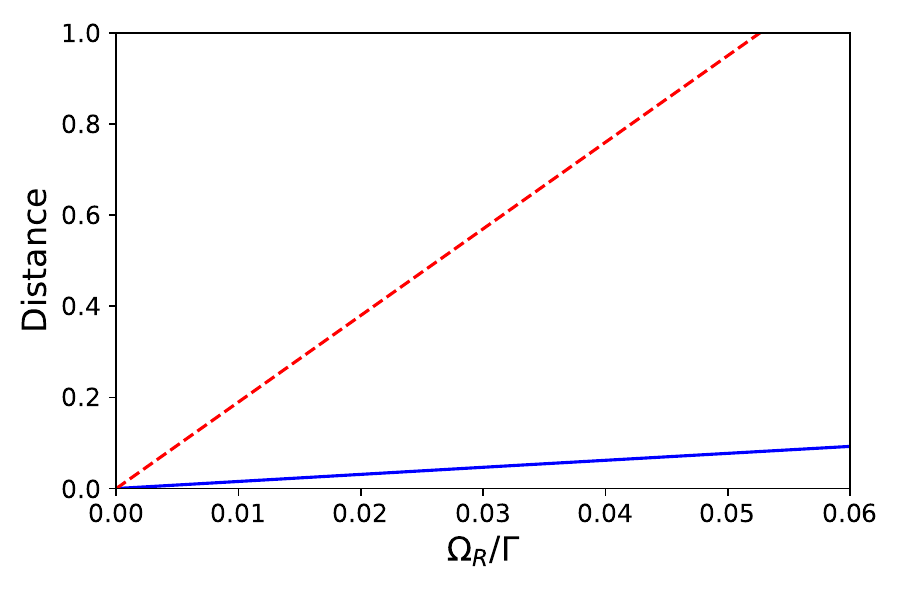}
    \caption{Distance between ideal and imperfect quantum Zeno gates as a function of $\frac{\Omega_R}{\Gamma}$. The solid blue line represents our metric, and the dashed red line depicts the bound by Blumenthal et al.}
    \label{fig:zeno_distance}
\end{figure}

Through our analysis, we identify that the imperfection of the quantum Zeno gate is due to the leakage of the doubly-excited state $\left|ee\right\rangle$ (refer to Appendix~\ref{app:Zeno} for a detailed analysis). Therefore, the distance defined in Equation~\ref{eq:Zeno} can be easily computed using the normalization distance given by Equation~\ref{eq:normalization_distance}. Figure~\ref{fig:zeno_distance} shows the distance between the ideal and imperfect quantum Zeno gates as a function of $\frac{\Omega_R}{\Gamma}$. The solid blue line represents the distance calculated by our metric, accurately quantifying the error. The dashed red line represents the ad-hoc metric ($d_\diamond<19\frac{\Omega_R}{\Gamma}$) from the original paper, which overestimates the error and is only applicable when $\frac{\Omega_R}{\Gamma}<0.06$. Our findings suggest potentially better performance for the quantum Zeno gate than previously anticipated, and our improved metric can benefit the design and engineering of these gates.

\section{Discussion and Conclusion}

Our proposed metric paves the way for the comprehensive analysis of general postselected protocols by accounting for measurement-induced disturbances. An ideal postselected protocol has a homogeneous measurement probability independent of the input state and reveals no information. However, this symmetry can be disrupted by errors, such as inaccurately prepared ancilla. In these cases, the measurements become biased and reveal information about the input state, introducing extra disturbance in addition to the original error associated with the ancilla. Consider, for example, the intricacies of quantum teleportation when using inaccurately prepared Bell pairs that are not maximally entangled as
\begin{equation*}
\left|B_{00}^\prime\right\rangle=\cos{\theta}\left|00\right\rangle+\sin{\theta}\left|11\right\rangle\;,
\end{equation*}
where $\theta\in\left(0,\pi/2\right)$ and deviates from the ideal value of $\pi/4$. Such deviation leads to input state-dependent measurement probabilities and teleportation channels, resulting in non-linear errors that conventional diamond distance cannot analyze. However, our metric offers a straightforward solution to quantify these errors, as expressed by
\begin{equation*}
d_g = \frac{\left|\cos{\theta}-\sin{\theta}\right|}{\cos{\theta}+\sin{\theta}}\;.
\end{equation*}
For those interested in a deeper analysis, the detailed calculations and additional non-trace-preserving error models for quantum teleportation and cluster-state quantum computing are presented in Appendix~\ref{app:examples}.

Our metric can also find important applications in fault-tolerant quantum computing. It provides a worst-case error bound for general error models and can be directly incorporated into the method of Aharonov et al.~\cite{Aharonov2008} to estimate fault-tolerant thresholds.
Additionally. the metric aids in analyzing fault tolerance through numerical simulations~\cite{Steane2003, Dawson2006a, Aliferis2006, Aliferis2008, Cross2009, Darmawan2017}, which offer a more precise approach for studying error propagation and calculating fault-tolerant thresholds. However, simulating general error models with classical computers becomes inefficient unless restricted to Clifford operators~\cite{Gottesman1997}. To address this issue, Magesan et al.~\cite{Magesan2013} proposed a method to approximate an error model using Clifford operators. They did this by solving an optimization problem, with the diamond distance serving as the objective function. However, the diamond distance only applies to trace-preserving operations. By employing our metric as the objective function instead of the diamond distance, we can approximate a non-trace-preserving error model using Clifford operators. This Clifford model enables the numerical study of how the error propagates through the quantum system.

In conclusion, we propose a metric for non-trace-preserving operations, characterizing their worst-case errors. This metric can be computed efficiently using renormalization techniques, offering a practical approach for quantifying non-trace-preserving errors and analyzing postselected protocols. As a result, the metric not only broadens the scope of imperfections that can be analyzed in quantum information processing, but also opens the door to a wide range of applications.

\begin{acknowledgments}

We would like to acknowledge Norbert Linke and Vladimir Manucharyan for their helpful discussions. Financial support was provided by the National Science Foundation under Grant Nos. OMA-2120757, OMA-1936314, and ECCS-1933546; the Air Force Office of Scientific Research under Grant Nos. UWSC12985 and FA23862014072; and the Army Research Laboratory under Grant No. W911NF1920181.

\end{acknowledgments}

\appendix

\section{Stability of the general distance\label{app:stability}}

We prove that the general distance, as defined in Equation~\ref{eq:general_distance}, satisfies stability, such that $d_g\left(\mathcal{E}\otimes\mathcal{I},\mathcal{F}\otimes\mathcal{I}\right)=d_g\left(\mathcal{E},\mathcal{F}\right)$ if $dim\left(X\right)\le dim\left(R\right)$, and it is optimized over a pure state on space $X\otimes R$. The proof is a generalization of Lemma 3.45 in~\cite{Watrous2018}. To begin, we first purify density operator $\rho$ by introducing an auxiliary space $R^\prime$, with $\rho=Tr_{R^\prime}\left[uu^\dag\right]$, where $u$ represents a pure state on space $X\otimes R\otimes R^\prime$. Therefore, Equation~\ref{eq:general_distance} can be reformulated as
\begin{widetext}
\begin{equation} \label{eq:kraus}
d_g\left(\mathcal{E},\mathcal{F}\right)=\frac{1}{2}\max_{u\in X\otimes R\otimes R^\prime}{\left\|Tr_{R^\prime}\left\{\frac{\sum_{i}{A_iuu^\dag A_i^\dag}}{Tr\left[\sum_{i}{A_iuu^\dag A_i^\dag}\right]}-\frac{\sum_{j}{B_juu^\dag B_j^\dag}}{Tr\left[\sum_{j}{B_juu^\dag B_j^\dag}\right]}\right\}\right\|_1}\;.
\end{equation}
\end{widetext}
In the above equation, $A_i$ and $B_j$ are Kraus operators of quantum operations $\mathcal{E}$ and $\mathcal{F}$, satisfying $\sum_{i}{A_i^\dag A_i}\le I$ and $\sum_{j}{B_j^\dag B_j}\le I$, respectively. We also abbreviate $A_i\otimes I$ as $A_i$ and $B_j\otimes I$ as $B_j$ for conciseness. Using the contractility of the trace distance~\cite{Nielsen2010}, $\frac{1}{2}\left\|Tr_{R^\prime}\left(\rho_1-\rho_2\right)\right\|_1\le\frac{1}{2}\left\|\rho_1-\rho_2\right\|_1$, Equation~\ref{eq:kraus} can be bounded by
\begin{widetext}
\begin{equation} \label{eq:contract}
d_g\left(\mathcal{E},\mathcal{F}\right)\le\frac{1}{2}\max_{u\in X\otimes R\otimes R^\prime}\left\|\frac{\sum_{i}{A_iuu^\dag A_i^\dag}}{\text{Tr}\left[\sum_{i}{A_iuu^\dag A_i^\dag}\right]}-\frac{\sum_{j}{B_juu^\dag B_j^\dag}}{\text{Tr}\left[\sum_{j}{B_juu^\dag B_j^\dag}\right]}\right\|_1\;.
\end{equation}
\end{widetext}
Even though Equation~\ref{eq:contract} appears as a non-strict inequality, we prove that equality always holds. By applying the Schmidt decomposition to the state $u$, we obtain 
$u=\sum_{l=1}^{n}{\sqrt{p_l}x_l\otimes z_l}$, where $p_l$ is the Schmidt coefficient, $x_l$ and $z_l$ are bases on spaces $X$ and $R\otimes R^\prime$, respectively, and $n$ is the dimension of space $X$. If $dim\left(X\right)\le dim\left(R\right)$, we can select a pure state $v\in X\otimes R$, such that $v=\sum_{l=1}^{n}{\sqrt{p_l}x_l\otimes e_l}$, where $e_l$ represents a basis on space $R$.
Defining an isometry $U=\sum_{l=1}^{n}{z_le_l^\dag}$, which maps from space $R$ to space $R \otimes R^\prime$, we obtain $u=\left(I_X\otimes U\right)v$, and derive the following identities
\begin{widetext}
\begin{eqnarray}
    \frac{\sum_{i}{A_iuu^\dag A_i^\dag}}{Tr\left[\sum_{i}{A_iuu^\dag A_i^\dag}\right]}&=&\frac{\sum_{i}{\left(A_i\otimes U\right)vv^\dag\left(A_i^\dag\otimes U^\dag\right)}}{Tr\left[\sum_{i}{\left(A_i\otimes U\right)vv^\dag\left(A_i^\dag\otimes U^\dag\right)}\right]}\nonumber\\
    &=&\frac{\sum_{i}{\left(I\otimes U\right)A_ivv^\dag A_i^\dag\left(I\otimes U^\dag\right)}}{Tr\left[\sum_{i}{\left(I\otimes U\right)A_ivv^\dag A_i^\dag\left(I\otimes U^\dag\right)}\right]}\nonumber\\
    &=&\left(I\otimes U\right)\frac{\sum_{i}{A_ivv^\dag A_i^\dag}}{Tr\left[\sum_{i}{A_ivv^\dag A_i^\dag}\right]}\left(I\otimes U^\dag\right)\;,
\end{eqnarray}
\end{widetext}
and a similar identity holds for quantum operation \( \mathcal{F} \), as
\begin{widetext}
    \begin{equation}
    \frac{\sum_{j}{B_juu^\dag B_j^\dag}}{Tr\left[\sum_{j}{B_juu^\dag B_j^\dag}\right]}=\left(I\otimes U\right)\frac{\sum_{j}{B_jvv^\dag B_j^\dag}}{Tr\left[\sum_{j}{B_jvv^\dag B_j^\dag}\right]}\left(I\otimes U^\dag\right)\;.
\end{equation}
\end{widetext}
The trace distance remains invariant under isometry, Thus, from Equation~\ref{eq:contract}, we derive
\begin{widetext}
\begin{eqnarray} \label{eq:pure_state}
    d_g\left(\mathcal{E},\mathcal{F}\right)&\le&\frac{1}{2}\max_{u\in X\otimes R\otimes R^\prime}{\left\|\frac{\sum_{i}{A_iuu^\dag A_i^\dag}}{Tr\left[\sum_{i}{A_iuu^\dag A_i^\dag}\right]}-\frac{\sum_{j}{B_juu^\dag B_j^\dag}}{Tr\left[\sum_{j}{B_juu^\dag B_j^\dag}\right]}\right\|_1}\nonumber\\
    &=&\frac{1}{2}\max_{v\in X\otimes R}{\left\|\left(I\otimes U\right)\left\{\frac{\sum_{i}{A_ivv^\dag A_i^\dag}}{Tr\left[\sum_{i}{A_ivv^\dag A_i^\dag}\right]}-\frac{\sum_{j}{B_jvv^\dag B_j^\dag}}{Tr\left[\sum_{j}{B_jvv^\dag B_j^\dag}\right]}\right\}\left(I\otimes U^\dag\right)\right\|_1}\nonumber\\
    &=&\frac{1}{2}\max_{v\in X\otimes R}{\left\|\frac{\sum_{i}{A_ivv^\dag A_i^\dag}}{Tr\left[\sum_{i}{A_ivv^\dag A_i^\dag}\right]}-\frac{\sum_{j}{B_jvv^\dag B_j^\dag}}{Tr\left[\sum_{j}{B_jvv^\dag B_j^\dag}\right]}\right\|_1}\nonumber\\
    &=&\frac{1}{2}\max_{v\in X\otimes R}{\left\|\frac{\left(\mathcal{E}\otimes\mathcal{I}_R\right)\left(vv^\dag\right)}{Tr\left[\left(\mathcal{E}\otimes\mathcal{I}_R\right)\left(vv^\dag\right)\right]}-\frac{\left(\mathcal{F}\otimes\mathcal{I}_R\right)\left(vv^\dag\right)}{Tr\left[\left(\mathcal{F}\otimes\mathcal{I}_R\right)\left(vv^\dag\right)\right]}\right\|_1}\;.
\end{eqnarray}
\end{widetext}
By comparing Equation~\ref{eq:pure_state} with Equation~\ref{eq:general_distance}, we find that the pure state $vv^\dag$ resides within a subset of the domain of optimization variable $\rho$ on space $X\otimes R$. Consequently, we can conclude that the stability of the general distance holds if $dim\left(X\right)\le dim\left(R\right)$, and the distance is optimized over a pure state on space $X\otimes R$.

\section{Quantum operation renormalization\label{app:renormalization}}

By introducing a renormalized state $u=\frac{\hat{A}\psi}{\left|\hat{A}\psi\right|}$, we can reformulate the general distance defined in Equation~\ref{eq:isometry} as
\begin{widetext}
\begin{equation} \label{eq:inverse_sp}
d_g\left(\mathcal{E},\mathcal{F}\right)=\frac{1}{2}\max_u{\left\|Tr_Z\left\{uu^\dag-\frac{\left(\hat{B}\hat{A}^{-1}\right)uu^\dag\left(\hat{B}\hat{A}^{-1}\right)^\dag}{u^\dag\left(\hat{B}\hat{A}^{-1}\right)^\dag\hat{B}\hat{A}^{-1}u}\right\}\right\|_1}\;,
\end{equation}
\end{widetext}
where $\hat{A}^{-1}$ denotes the pseudoinverse of $\hat{A}$. Applying singular value decomposition to $\hat{B}\hat{A}^{-1}$, we obtain
\begin{equation*}
    \hat{B}\hat{A}^{-1}=UMV^\dag\;.
\end{equation*}
In this expression, $U$ and $V$ are isometries  that map the input space $X$ to the output space $Z\otimes Y$. Specifically, $U$ is of dimension $k^\prime n\times m$, $V$ is $kn\times m$, and $M$ is a nonnegative diagonal operator with a dimension of $m\times m$. Here, $m$ and $n$ denote the dimensions of spaces $X$ and $Y$, respectively, while $k$ and $k^\prime$ represent the cardinalities of the sets of Kraus operators $\{A_i\}$ and $\{B_j\}$. With this decomposition, we can write Equation~\ref{eq:inverse_sp} as
\begin{equation} \label{eq:svd_sp}
d_g\left(\mathcal{E},\mathcal{F}\right)=\frac{1}{2}\max_u{\left\|Tr_Z\left\{uu^\dag-\frac{UMV^\dag uu^\dag VMU^\dag}{u^\dag VM^2V^\dag u}\right\}\right\|_1}\;.
\end{equation}
By defining a new state $v=V^\dag u$ and using the triangular inequality and the contractility of trace distance, we can bound Equation~\ref{eq:svd_sp} by
\begin{widetext}
\begin{eqnarray} \label{eq:derive_bound_sp}
    d_g\left(\mathcal{E},\mathcal{F}\right)&=&\frac{1}{2}\max_v{\left\|Tr_Z\left\{V vv^\dag V^\dag-\frac{UM vv^\dag M U^\dag}{v^\dag M^2v}\right\}\right\|_1}\nonumber\\
    &=&\frac{1}{2}\max_v{\left\|Tr_Z\left\{V vv^\dag V^\dag-U vv^\dag U^\dag+U vv^\dag U^\dag-\frac{UM vv^\dag M U^\dag}{v^\dag M^2v}\right\}\right\|_1}\nonumber\\
    &\le&\frac{1}{2}\max_v{\left\|Tr_Z\left\{Vvv^\dag V^\dag-Uvv^\dag U^\dag\right\}\right\|_1}+\frac{1}{2}\max_v{\left\|Tr_Z\left\{Uvv^\dag U^\dag-\frac{UMvv^\dag M U^\dag}{v^\dag M^2v}\right\}\right\|_1}\nonumber\\
    &\le&\frac{1}{2}\max_v{\left\|Tr_Z\left\{Vvv^\dag V^\dag-Uvv^\dag U^\dag\right\}\right\|_1}+\frac{1}{2}\max_v{\left\|vv^\dag-\frac{Mvv^\dag M}{v^\dag M^2v}\right\|_1}\;.
\end{eqnarray}
\end{widetext}
We define two renormalized quantum channels
\begin{eqnarray}
    \mathcal{U}\left(\rho\right)&=&Tr_Z\left\{U\rho U^\dag\right\}\;,\nonumber\\
    \mathcal{V}\left(\rho\right)&=&Tr_Z\left\{V\rho V^\dag\right\}\;,
\end{eqnarray}
and a normalization operation
\begin{equation}
    \mathcal{M}\left(\rho\right)=\frac{M\rho M}{Tr\left[M\rho M\right]}\;.
\end{equation}
The general distance between $\mathcal{E}$ and $\mathcal{F}$ is bounded by Equation~\ref{eq:derive_bound_sp}, as
\begin{equation}
    d_g\left(\mathcal{E},\mathcal{F}\right)\le d_\diamond\left(\mathcal{U},\mathcal{V}\right)+d_g\left(\mathcal{I},\mathcal{M}\right)\;,
\end{equation}
where $d_\diamond\left(\mathcal{U},\mathcal{V}\right)$ denotes the conventional diamond distance, which can be computed using convex optimization; $d_g\left(\mathcal{I},\mathcal{M}\right)$ represents the general distance between an identity operation and the normalization operation, which we refer to as the normalization distance.

\section{Calculation of the normalization distance\label{app:normalization_distance}}

We can directly calculate the normalization distance
\begin{equation}
    d_g\left(\mathcal{I},\mathcal{M}\right)=\frac{1}{2}\max_\psi{\left\|\psi\psi^\dag-\frac{M\psi\psi^\dag M}{\left\langle\psi\middle|M^2\middle|\psi\right\rangle}\right\|_1}\;,
\end{equation}
where $\psi$ is a pure quantum state, and $M$ is a nonnegative diagonal operator with a dimension of $n\times n$. We can denote the normalized output state as $\left|\phi\right\rangle=\frac{M\left|\psi\right\rangle}{\left|M\left|\psi\right\rangle\right|}$, and the normalization distance is given by
\begin{equation}
    d_g\left(\mathcal{I},\mathcal{M}\right)=\frac{1}{2}\max_\psi{\left\|\psi\psi^\dag-\phi\phi^\dag\right\|_1}\;.
\end{equation}
The trace distance between two pure states is related to their inner product by 
\begin{equation*}
\frac{1}{2}\left\|\psi\psi^\dag-\phi\phi^\dag\right\|_1=\sqrt{1-\left|\left\langle\psi\middle|\phi\right\rangle\right|^2}\;,
\end{equation*}
so that maximizing $\left\|\psi\psi^\dag-\phi\phi^\dag\right\|_1$ is equivalent to minimizing $\cos{\theta}=\left|\left\langle\psi\middle|\phi\right\rangle\right|$, where we refer to $\theta$ as the rotating angle between the input and output states.

We can compute $\cos{\theta}$ by induction. We denote the normalization operator as
\begin{equation*}
M=diag\left(\lambda_1,\lambda_2,\cdots,\lambda_n\right)\;,
\end{equation*}
where $\lambda_1\ge\lambda_2\ge\cdots\ge\lambda_n\ge0$. We first consider the projection of $M$ on a $l$-dimensional subspace as
\begin{equation*}
M_l=diag\left(\lambda_1,\lambda_2,\cdots,\lambda_l\right)\;.
\end{equation*}
A vector $\left|u\right\rangle$ on the subspace can be decomposed as
\begin{equation*}
\left|u\right\rangle=\sqrt x\left|l\right\rangle+\sqrt{1-x}\left|v\right\rangle\;,
\end{equation*}
where $x\in\left[0,1\right]$, $\left|l\right\rangle$ is the basis of $M$ corresponding to the eigenvalue $\lambda_l$, and $\left|v\right\rangle$ is a unit vector on the $\left(l-1\right)$-dimensional subspace orthogonal to $\left|l\right\rangle$. The assumption of positive coefficients loses no generality because we can always redefine the basis with any phase. Applying $M_l$ to $\left|u\right\rangle$ gives
\begin{equation}
    M_l\left|u\right\rangle=\lambda_l\sqrt x\left|l\right\rangle+\sqrt{1-x}M_{l-1}\left|v\right\rangle\;,
    \label{eq:ml_sp}
\end{equation}
where $M_{l-1}$ is the projection of $M$ on the $\left(l-1\right)$-dimensional subspace. Applying $M_{l-1}$ to $\left|v\right\rangle$ results
\begin{equation}
    M_{l-1}\left|v\right\rangle=r_{l-1}\cos{\theta_{l-1}}\left|v\right\rangle+r_{l-1}\sin{\theta_{l-1}}\left|v_\perp\right\rangle\;,
    \label{eq:ml1_sp}
\end{equation}
where $r_{l-1}$ and $\theta_{l-1}$ are the contracting factor and rotating angle, respectively, and $\left|v_\perp\right\rangle$ represents a unit vector orthogonal to $\left|v\right\rangle$. A specific rotating axis does not affect the following calculation because the axis is always orthogonal to $\left|l\right\rangle$. The contracting factor and rotating angle of $M_l\left|u\right\rangle$ are given by $r_l=\sqrt{\left|\left\langle u\middle|M_l^2\middle|u\right\rangle\right|}$ and $\cos{\theta_l}=\left\langle u\middle|M_l\middle|u\right\rangle/r_l$, respectively, and combining Equations~\ref{eq:ml_sp} and~\ref{eq:ml1_sp} gives
\begin{eqnarray}
    r_l^2&=&x\lambda_l^2+\left(1-x\right)r_{l-1}^2\;,\nonumber\\
    r_l\cos{\theta_l}&=&x\lambda_l+\left(1-x\right)r_{l-1}\cos{\theta_{l-1}}\;,
\end{eqnarray}
i.e., $\left(r_l^2,r_l\cos{\theta_l}\right)$ is a convex combination of $\left(\lambda_l^2,\lambda_l\right)$ and $\left(r_{l-1}^2,r_{l-1}\cos{\theta_{l-1}}\right)$. By induction, the feasible domain of $\left(r^2,r\cos{\theta}\right)$ for the normalization operation $\mathcal{M}$ is given by
\begin{eqnarray}
    r^2&=&\sum_{i}{x_i\lambda_i^2}\nonumber\\
    r\cos{\theta}&=&\sum_{i}{x_i\lambda_i}\;,
    \label{eq:convex_sp}
\end{eqnarray}
where $x_i$ represents the coefficient of convex combination satisfying $\sum_{i}{x_i}=1$, and $\lambda_i$ is the eigenvalue of $M$. The minimum of $\cos{\theta}$ is on the boundary of the feasible domain depending on the combination two eigenvalues. We can calculate the minimum of $\cos{\theta}$ easily by calculus, as
\begin{equation}
    \min{\cos{\theta}}=\frac{2\sqrt{\lambda_{\rm max}\lambda_{\rm min}}}{\lambda_{\rm max}+\lambda_{\rm min}}\;,
\end{equation}
where $\lambda_{\rm max}$ and $\lambda_{\rm min}$ are the maximum and minimum eigenvalues of $M$. Therefore, the distance between the identity and normalization operations is given by
\begin{equation}
    d_g\left(\mathcal{I},\mathcal{M}\right)=\frac{\lambda_{\rm max}-\lambda_{\rm min}}{\lambda_{\rm max}+\lambda_{\rm min}}\;.
\end{equation}

\section{Examples of non-trace-preserving errors in postselected protocols\label{app:examples}}

\subsection{Inaccurate Bell pairs in quantum teleportation}

The inaccurately prepared Bell pairs can cause non-trace-preserving errors in quantum teleportation, which has been discussed for trapped-ions~\cite{Moehring2007a} and single rail photonic qubits~\cite{Dhara2023}. We assume that other operations, including Bell measurements and single-qubit rotations, are ideal, but the prepared Bell pairs are in state
\begin{equation*}
\left|B_{00}^\prime\right\rangle=\cos{\theta}\left|00\right\rangle+\sin{\theta}\left|11\right\rangle\;,
\end{equation*}
where $\theta\in\left(0,\frac{\pi}{2}\right)$ and deviates from the ideal value $\frac{\pi}{4}$. To derive imperfect channels, we assume that the input qubit is in state $\left|\psi\right\rangle=\alpha\left|0\right\rangle+\beta\left|1\right\rangle$. We then represent the first two qubits of the total state $\left|\psi\right\rangle\otimes\left|B_{00}^\prime\right\rangle$ in Bell bases and obtain
\begin{eqnarray*}
    \left|\psi\right\rangle\otimes\left|B_{00}^\prime\right\rangle&=&\left|B_{00}\right\rangle\otimes\left(\frac{\alpha\cos{\theta}}{\sqrt2}\left|0\right\rangle+\frac{\beta\sin{\theta}}{\sqrt2}\left|1\right\rangle\right)\\
    & &+\left|B_{01}\right\rangle\otimes\left(\frac{\alpha\cos{\theta}}{\sqrt2}\left|0\right\rangle-\frac{\beta\sin{\theta}}{\sqrt2}\left|1\right\rangle\right)\\
    & &+\left|B_{10}\right\rangle\otimes\left(\frac{\alpha\sin{\theta}}{\sqrt2}\left|1\right\rangle+\frac{\beta\cos{\theta}}{\sqrt2}\left|0\right\rangle\right)\\
    & &+\left|B_{00}\right\rangle\otimes\left(\frac{\alpha\sin{\theta}}{\sqrt2}\left|1\right\rangle-\frac{\beta\cos{\theta}}{\sqrt2}\left|0\right\rangle\right)\;,
\end{eqnarray*}
where $\left|B_{\rm mn}\right\rangle$ ($\rm mn$ is a binary number) is the Bell basis and given by
\begin{eqnarray}
    \left|B_{00}\right\rangle&=&\frac{\sqrt2}{2}\left(\left|00\right\rangle+\left|11\right\rangle\right)\;,\nonumber\\
    \left|B_{01}\right\rangle&=&\frac{\sqrt2}{2}\left(\left|00\right\rangle-\left|11\right\rangle\right)\;,\nonumber\\
    \left|B_{10}\right\rangle&=&\frac{\sqrt2}{2}\left(\left|01\right\rangle+\left|10\right\rangle\right)\;,\nonumber\\
    \left|B_{11}\right\rangle&=&\frac{\sqrt2}{2}\left(\left|01\right\rangle-\left|10\right\rangle\right)\;.
\end{eqnarray}
Therefore, the quantum operations corresponding to each Bell measurement is given by
\begin{equation}
    \mathcal{E}_{\rm mn}\left(\rho\right)=\frac{E_{\rm mn}\rho E_{\rm mn}^\dag}{Tr\left[E_{\rm mn}\rho E_{\rm mn}^\dag\right]}\;,
\end{equation}
where the corresponding Kraus operators are given by
\begin{eqnarray*}
    &E_{00}=\begin{pmatrix}\frac{\cos{\theta}}{\sqrt2}&0\\0&\frac{\sin{\theta}}{\sqrt2}\end{pmatrix}\;,\quad &E_{01}=\begin{pmatrix}\frac{\cos{\theta}}{\sqrt2}&0\\0&-\frac{\sin{\theta}}{\sqrt2}\end{pmatrix}\;,\\
    &E_{10}=\begin{pmatrix}0&\frac{\cos{\theta}}{\sqrt2}\\\frac{\sin{\theta}}{\sqrt2}&0\end{pmatrix}\;,\quad &E_{11}=\begin{pmatrix}0&-\frac{\cos{\theta}}{\sqrt2}\\\frac{\sin{\theta}}{\sqrt2}&0\end{pmatrix}\;.
\end{eqnarray*}
We can compute the general distance between channel $\mathcal{E}_{\rm mn}$ and unitary $U_{\rm mn}\in\{I,Z,X,XZ\}$. They result in the same distance, given by
\begin{equation}
    d_g\left(\mathcal{E}_{\rm mn},U_{\rm mn}\right)=\frac{\left|\cos{\theta}-\sin{\theta}\right|}{\cos{\theta}+\sin{\theta}}\;.
\end{equation}

\subsection{Probabilistic CNOT errors in quantum teleportation}

Probabilistic $CNOT$ errors on Bell pairs can cause non-trace-preserving errors in quantum teleportation. We consider a channel that applies $CONT$ gates on the ideal Bell pair $\left|B_{00}\right\rangle$ with probability $p$, given by
\begin{widetext}
\begin{equation}
    \mathcal{E}\left(\left|B_{00}\middle\rangle\middle\langle B_{00}\right|\right)=\left(1-p\right)\left|B_{00}\middle\rangle\middle\langle B_{00}\right|+p{CNOT}\left|B_{00}\middle\rangle\middle\langle B_{00}\right|{CNOT}^\dag\;.
\end{equation}
\end{widetext}
This channel can model the non-deterministic generation of Bell pairs, but the success event is not perfectly heralded. We can represent the resulting mixed state $\Phi_{00}=\mathcal{E}\left(\left|B_{00}\middle\rangle\middle\langle B_{00}\right|\right)$ by state purification, denoted as
\begin{widetext}
\begin{equation}
    \left|\Phi_{00}\right\rangle=\sqrt{\frac{1-p}{2}}\left(\left|00\right\rangle+\left|11\right\rangle\right)\otimes\left|e_0\right\rangle+\sqrt{\frac{p}{2}}\left(\left|00\right\rangle+\left|10\right\rangle\right)\otimes\left|e_1\right\rangle\;.
\end{equation}
\end{widetext}
To derive imperfect channels, we assume an input qubit in state $\left|\psi\right\rangle=\alpha\left|0\right\rangle+\beta\left|1\right\rangle$. We then represent the first two qubits of the total state $\left|\psi\right\rangle\otimes\left|\Phi_{00}\right\rangle$ in Bell bases and obtain
\begin{widetext}
\begin{eqnarray}
    \left|\psi\right\rangle\otimes\left|\Phi_{00}\right\rangle&=&\left|B_{00}\right\rangle\otimes\left[\frac{\sqrt{1-p}}{2}\left(\alpha\left|0\right\rangle+\beta\left|1\right\rangle\right)\otimes\left|e_0\right\rangle+\frac{\sqrt p}{2}\left(\alpha+\beta\right)\left|0\right\rangle\otimes\left|e_1\right\rangle\right]\nonumber\\
    & &+\left|B_{01}\right\rangle\otimes\left[\frac{\sqrt{1-p}}{2}\left(\alpha\left|0\right\rangle-\beta\left|1\right\rangle\right)\otimes\left|e_0\right\rangle+\frac{\sqrt p}{2}\left(\alpha-\beta\right)\left|0\right\rangle\otimes\left|e_1\right\rangle\right]\nonumber\\
    & &+\left|B_{10}\right\rangle\otimes\left[\frac{\sqrt{1-p}}{2}\left(\alpha\left|1\right\rangle+\beta\left|0\right\rangle\right)\otimes\left|e_0\right\rangle+\frac{\sqrt p}{2}\left(\alpha+\beta\right)\left|0\right\rangle\otimes\left|e_1\right\rangle\right]\nonumber\\
    & &+\left|B_{11}\right\rangle\otimes\left[\frac{\sqrt{1-p}}{2}\left(\alpha\left|1\right\rangle-\beta\left|0\right\rangle\right)\otimes\left|e_0\right\rangle+\frac{\sqrt p}{2}\left(\alpha+\beta\right)\left|0\right\rangle\otimes\left|e_1\right\rangle\right]\;.
\end{eqnarray}
\end{widetext}
Assuming that the Bell measurement gives a result of $\left|B_{00}\right\rangle$, the operation on an arbitrary input state $\rho$ is given by
\begin{equation}
    \mathcal{E}_{00}\left(\rho\right)=\frac{E_0\rho E_0^\dag+E_1\rho E_1^\dag}{Tr\left[E_0\rho E_0^\dag+E_1\rho E_1^\dag\right]}\;,
\end{equation}
where the Kraus operators are given by
\begin{eqnarray*}
    E_0=\begin{pmatrix}\frac{\sqrt{1-p}}{2}&0\\0&\frac{\sqrt{1-p}}{2}\end{pmatrix}\;,\quad E_1=\begin{pmatrix}\frac{\sqrt p}{2}&\frac{\sqrt p}{2}\\0&0\end{pmatrix}\;.
\end{eqnarray*}
The operation is non-linear, and its probabilities can be characterized by a semidefinite operator $S=E_0^\dag E_0+E_1^\dag E_1$~\cite{Bongioanni2010}. Applying the singular value decomposition to $S$ results in
\begin{equation}
    S=H\begin{pmatrix}\frac{1+p}{4}&0\\0&\frac{1-p}{4}\end{pmatrix}H^\dag\;,
\end{equation}
where $H$ is the Hadamard gate. The singular values represent the measurement probabilities for different state bases. It shows that the channel has a state-dependent measurement probability, which will have a non-linear effect on the teleported state. Therefore, we should use the general distance to quantify its error and denote it as $d_g\left(\mathcal{E}_{00},I\right)$. We did not obtain a closed-form of the distance, but it can be numerically bounded by our algorithm. We omit the analysis of the other Bell measurements. They follow the same analysis steps and yield the same distances.

\subsection{Amplitude damping errors in cluster-state quantum computing}
To show non-trace-preserving errors caused by amplitude damping in cluster-state quantum computing, we consider a measurement-based Hadamard gate. To implement a Hadamard gate on state $\left|\psi\right\rangle=\alpha\left|0\right\rangle+\beta\left|1\right\rangle$, as shown in the circuit
\[
\Qcircuit @C=1em @R=1em @!R {
 \lstick{\left|\psi\right\rangle} & \ctrl{1}     & \gate{H} & \meter \\
 \lstick{\left|+\right\rangle}    & \control \qw & \qw      & \rstick{X^m H\left|\psi\right\rangle}
 \qw
}
\]
we apply a controlled-Z gate between the input state and an ancilla state $\left|+\right\rangle$, and measure the first qubit in bases $\left\{\left|+\right\rangle,\left|-\right\rangle\right\}$. The output state is encoded in the second qubit, which goes through a Hadamard gate with a conditional Pauli-X gate depending on the measurement result. Assume that the ancilla $\left|+\right\rangle$ suffers from an amplitude damping channel, given by
\begin{equation}
    \mathcal{E}_{\rm AD}\left(\left|+\middle\rangle\middle\langle+\right|\right)=E_0\left|+\middle\rangle\middle\langle+\right|E_0^\dag+E_1\left|+\middle\rangle\middle\langle+\right|E_1^\dag\;,
\end{equation}
where $E_0=\begin{pmatrix}1&0\\0&\sqrt{1-p}\end{pmatrix}$ and $E_1=\begin{pmatrix}0&\sqrt p\\0&0\end{pmatrix}$, and $p$ is the damping probability. We can represent the imperfect ancilla $\Phi_{AD}=\mathcal{E}_{\rm AD}\left(\left|+\middle\rangle\middle\langle+\right|\right)$ by state purification, denoted as
\begin{equation*}
    \left|\Phi_{\rm AD}\right\rangle=\frac{1}{\sqrt2}\left(\left|0\right\rangle+\sqrt{1-p}\left|1\right\rangle\right)\otimes\left|e_0\right\rangle+\sqrt{\frac{p}{2}}\left|0\right\rangle\otimes\left|e_1\right\rangle\;.
\end{equation*}
After a controlled-Z gate and a Hadamard gate on the total state $\left|\psi\right\rangle\otimes\left|\Phi_{\rm AD}\right\rangle$, the resulting state is given by
\begin{widetext}
\begin{eqnarray}
    \left|\psi\right\rangle\otimes\left|\Phi_{\rm AD}\right\rangle&\rightarrow&\left|0\right\rangle\otimes\left[\frac{\alpha+\beta}{2}\left|0\right\rangle+\frac{\left(\alpha-\beta\right)\sqrt{1-p}}{2}\left|1\right\rangle\right]\otimes\left|e_0\right\rangle\nonumber\\
    & &+\left|0\right\rangle\otimes\frac{\left(\alpha+\beta\right)\sqrt p}{2}\left|0\right\rangle\otimes\left|e_1\right\rangle\nonumber\\
    & &+\left|1\right\rangle\otimes\left[\frac{\alpha-\beta}{2}\left|0\right\rangle+\frac{\left(\alpha+\beta\right)\sqrt{1-p}}{2}\left|1\right\rangle\right]\otimes\left|e_0\right\rangle\nonumber\\
    & &+\left|1\right\rangle\otimes\frac{\left(\alpha-\beta\right)\sqrt p}{2}\left|0\right\rangle\otimes\left|e_1\right\rangle\;.
\end{eqnarray}
\end{widetext}
Assuming that the measurement of the first qubit is $\left|0\right\rangle$, the imperfect Hadamard gate is given by
\begin{equation}
    \mathcal{H}_{\rm AD}\left(\rho\right)=\frac{F_0\rho F_0^\dag+F_1\rho F_1^\dag}{Tr\left[F_0\rho F_0^\dag+F_1\rho F_1^\dag\right]}\;,
\end{equation}
where the Kraus operators are given by
\begin{eqnarray*}
    F_0=\begin{pmatrix}\frac{1}{2}&\frac{1}{2}\\\frac{\sqrt{1-p}}{2}&-\frac{\sqrt{1-p}}{2}\end{pmatrix}\;,\quad
    F_1=\begin{pmatrix}\frac{\sqrt p}{2}&\frac{\sqrt p}{2}\\0&0\end{pmatrix}\;.
\end{eqnarray*}
The operation has a state-dependent probability, characterized by a semidefinite operator $S=F_0^\dag F_0+F_1^\dag F_1$. Applying the singular value decomposition to $S$ results in
\begin{equation}
    S=H\begin{pmatrix}\frac{1+p}{2}&0\\0&\frac{1-p}{2}\end{pmatrix}H^\dag\;,
\end{equation}
where $H$ is the Hadamard gate. The singular values represent the measurement probabilities for different state bases. We can quantify the error of the imperfect Hadamard operation by the general distance $d_g\left(\mathcal{H}_{\rm AD},H\right)$. Our metric calculates an upper bound on the distance as
\begin{equation}
    d_g\left(\mathcal{H}_{\rm AD},H\right)\le\frac{\sqrt{1+p}-1}{2\sqrt{1+p}}+\frac{1-\sqrt{1-p^2}}{p}\;.
\end{equation}
We omit the analysis of measurement $\left|1\right\rangle$, which follows the same analysis steps and yield the same metric.

\section{Revisiting the Pauli twirling approximation\label{app:PTA}}

In this appendix, we revisit the validity of the Pauli twirling approximation for the no-leak channel of neutral atoms in~\cite{Wu2022}. The postselected channel with perfect detection efficiency is given by
\begin{equation}
    \mathcal{E}_K\left(\rho\right)=\frac{K\rho K^\dag}{Tr\left[K\rho K^\dag\right]}\;,
\end{equation}
where $K=\left|0\middle\rangle\middle\langle0\right|+\sqrt{1-p}\left|1\middle\rangle\middle\langle\right|$ is the no-leak operator, and $p$ is the leakage probability. Wu et al. analyzed the channel by applying the Pauli twirling approximation~\cite{Silva2008} to operator $K$, which can be done by removing the off-diagonal elements in the process matrix of operation $K\rho K^\dag$ represented in Pauli bases. Operator $K$ is denoted by 
\begin{equation*}
K=\frac{1+\sqrt{1-p}}{2}I+\frac{1-\sqrt{1-p}}{2}Z\;,
\end{equation*}
so the operation is given by
\begin{widetext}
\begin{equation}
    K\rho K^\dag=\frac{\left(1+\sqrt{1-p}\right)^2}{4}\rho+\frac{p}{4}Z\rho+\frac{p}{4}\rho Z^\dag+\frac{\left(1-\sqrt{1-p}\right)^2}{4}Z\rho Z^\dag\;.
\end{equation}
\end{widetext}
By removing the off-diagonal terms of the operation, we obtain its Pauli twirling approximation
\begin{equation}
    \left.K\rho K^\dag\right|_{\rm PTA}=\frac{\left(1+\sqrt{1-p}\right)^2}{4}\rho+\frac{\left(1-\sqrt{1-p}\right)^2}{4}Z\rho Z^\dag\;.
    \label{eq:unnormalized_PTA}
\end{equation}
However, this approximation is still non-trace-preserving, so we should normalize it to get
\begin{widetext}
\begin{eqnarray}
    \mathcal{E}_K^\prime\left(\rho\right)&=&\frac{\left.K\rho K^\dag\right|_{\rm PTA}}{Tr\left[\left.K\rho K^\dag\right|_{\rm PTA}\right]}\nonumber\\
    &=&\frac{\left(1+\sqrt{1-p}\right)^2}{4-2p}\rho+\frac{\left(1-\sqrt{1-p}\right)^2}{4-2p}Z\rho Z^\dag\;.
    \label{eq:PTA}
\end{eqnarray}
\end{widetext}
Then, the diamond distance between the approximated and identity channels can quantify the error strength as
\begin{equation}
    d_\diamond\left(\mathcal{E}_K^\prime,I\right)=\frac{\left(1-\sqrt{1-p}\right)^2}{4-2p}\;.
\end{equation}
Wu et al. argued that when $p$ is small, the error strength behaves as $\frac{P^2}{16}$ and is therefore negligible.

However, we can directly calculate the general distance between the postselected and identity channels by Equation 8 in the main text, namely $d_g=\frac{\lambda_{\rm max}-\lambda_{\rm min}}{\lambda_{\rm max}+\lambda_{\rm min}}$. By inserting $\lambda_{\rm max}=1$ and $\lambda_{\rm min}=\sqrt{1-p}$ into the equation, we obtain the distance
\begin{equation}
    d_g\left(\mathcal{E},I\right)=\frac{1-\sqrt{1-p}}{1+\sqrt{1-p}}\;.
    \label{eq:leakage_distance}
\end{equation}
It behaves as $\frac{p}{4}$ when $p$ is small. In contrast, the Pauli twirling approximation underestimates the errors because it does not account for the state-dependent renormalization effect that can lead to extra errors. 

We note that the diamond distance between the unnormalized approximated channel (Equation~\ref{eq:unnormalized_PTA}) and the identity channel is given by
\begin{equation}
    d_\diamond\left(\left.K\rho K^\dag\right|_{\rm PTA},I\right)=\frac{1-\sqrt{1-p}}{2}\;.
\end{equation}
When $p$ is small, it approaches the actual error strength in Equation~\ref{eq:leakage_distance}. However, the unnormalized channel lacks a physical interpretation. We also note that Wu et al. considered the behavior of the postselected no-leak channel and an amplitude damping channel to be similar. However, the amplitude damping channel is trace-preserving and has an additional damping operator $J=\sqrt{p}\left|0\middle\rangle\middle\langle1\right|$. Its complete representation is given by $\mathcal{E}_A=K\rho K^\dag+J\rho J^\dag$. The diamond distance between the amplitude damping channel and the identity channel is given by $d_\diamond\left(\mathcal{E}_A,I\right)=p$. The Pauli twirling approximation for the amplitude damping channel is given by
\begin{widetext}
\begin{equation}
    \mathcal{E}_A^\prime\left(\rho\right)=\frac{\left(1+\sqrt{1-p}\right)^2}{4}\rho+\frac{p}{4}X\rho X^\dag+\frac{p}{4}Y\rho Y^\dag+\frac{\left(1-\sqrt{1-p}\right)^2}{4}Z\rho Z^\dag\;.
\end{equation}
\end{widetext}
It also applies the Pauli-X and Pauli-Y errors at a rate of $\frac{p}{4}$ compared to the Pauli twirling approximation for the no-leak channel in Equation~\ref{eq:PTA}.

\section{Calculating the distance of leakage channels\label{app:leakge_reduction}}

In this section, we quantify the errors of neutral atoms under leakage detection. The undetected leakage channel is given by
\begin{equation}
    \mathcal{E}_{\rm leak}\left(\rho\right)=K\rho K^\dag+L\rho L^\dag\;.
\end{equation}
In the above equation, $K=\left|0\middle\rangle\middle\langle0\right|+\left(1-\sqrt{1-p}\right)\left|1\middle\rangle\middle\langle1\right|$ is the Kraus operator corresponding to a no-leak evolution, and $L=\sqrt{p}\left|e\middle\rangle\middle\langle1\right|$ is the Kraus operator corresponding to the leakage, where $p$ is the leakage probability, $\left|0\right\rangle$ and $\left|1\right\rangle$ are the computational states, and $\left|e\right\rangle$ is an irrelevant state outside the computational subspace. The error between the undetected leakage channel and the identity channel can be measured by the diamond distance, which is given by
\begin{equation}
    d_\diamond\left(\mathcal{E}_{\rm leak},I\right)=p\;.
\end{equation}

In the case of imperfect detection efficiency, the postselected channel under leakage detection is given by
\begin{equation}
    \mathcal{E}_{\rm post}\left(\rho\right)=\frac{K\rho K^\dag+\left(1-\gamma\right)L\rho L^\dag}{Tr\left[K\rho K^\dag+\left(1-\gamma\right)L\rho L^\dag\right]}\;,
\end{equation}
where $\gamma$ is the detection efficiency. Since the channel is non-trace-preserving, we should quantify its error by the general distance $d_g\left(\mathcal{E}_{\rm post},I\right)$. Using the renormalization technique, we can bound the general distance by
\begin{equation}
    d_g\left(\mathcal{E}_{\rm post},I\right)\le d_\diamond\left(\mathcal{U},I\right)+d_g\left(\mathcal{M},I\right)\;,
\end{equation}
where $\mathcal{U}$ is a trace-preserving operation, and $\mathcal{M}$ is a normalizing operation. In the operator-sum representation, operation $\mathcal{U}$ is given by
\begin{equation}
    \mathcal{U}\left(\rho\right)=E_0\rho E_0^\dag+E_1\rho E_1^\dag\;,
\end{equation}
where
\begin{eqnarray*}
    E_0=\begin{pmatrix}1&0\\0&\sqrt{\frac{1-p}{1-\gamma p}}\\0&0\end{pmatrix}\;,\quad
    E_1=\begin{pmatrix}0&0\\0&0\\0&\sqrt{\frac{\left(1-\gamma\right)p}{1-\gamma p}}\end{pmatrix}\;,
\end{eqnarray*}
representing maps from input space $\left\{\left|0\right\rangle,\left|1\right\rangle\right\}$ to output space $\left\{\left|0\right\rangle,\left|1\right\rangle,\left|e\right\rangle\right\}$. We can calculate the diamond distance as
\begin{equation}
    d_\diamond\left(\mathcal{U},I\right)=\frac{\left(1-\gamma\right)p}{1-\gamma p}\;.
\end{equation}
The normalizing operation $\mathcal{M}$ is represented by
\begin{equation}
    \mathcal{M}\left(\rho\right)=\frac{M\rho M^\dag}{Tr\left[M\rho M^\dag\right]}\;,
\end{equation}
where $M=\begin{pmatrix}1&0\\0&\sqrt{1-\gamma p}\end{pmatrix}$. Its distance from the identity channel is given by
\begin{equation}
    d_g\left(\mathcal{M},I\right)=\frac{1-\sqrt{1-\gamma p}}{1+\sqrt{1-\gamma p}}\;.
\end{equation}
Therefore, the general distance between the detected leakage channel and the identity channel is bounded by
\begin{equation}
    d_g\left(\mathcal{E}_{\rm post},I\right)\le\frac{\left(1-\gamma\right)p}{1-\gamma p}+\frac{1-\sqrt{1-\gamma p}}{1+\sqrt{1-\gamma p}}\;.
\end{equation}
In particular, when one has perfect detection ($\gamma=1$), the bound is exact, and the general distance is given by $\frac{1-\sqrt{1-p}}{1+\sqrt{1-p}}$.

\section{Quantum Zeno gate\label{app:Zeno}}

The exponent of the imperfect Zeno transformation in Equation~\ref{eq:Zeno_transformation} is given by
\begin{widetext}
\begin{equation}
    t\left(\Gamma\mathcal{D}-i\mathcal{H}\right)=\pi\left(\left|eg\middle\rangle\middle\langle fg\right|-\left|fg\middle\rangle\middle\langle eg\right|+\left|ee\middle\rangle\middle\langle fe\right|-\left|fe\middle\rangle\middle\langle ee\right|-\frac{1}{r}\left|fe\middle\rangle\middle\langle fe\right|\right)\;,
\end{equation}
\end{widetext}
where we relabel $r=\frac{\Omega_R}{\Gamma}$. The transformation is calculated by matrix exponentiation, resulting in
\begin{widetext}
\begin{equation}
    P_0e^{t\left(\Gamma\mathcal{D}-i\mathcal{H}\right)}P_0=\left|gg\middle\rangle\middle\langle gg\right|+\left|ge\middle\rangle\middle\langle ge\right|-\left|eg\middle\rangle\middle\langle eg\right|+\lambda\left|ee\middle\rangle\middle\langle ee\right|\;,
\end{equation}
\end{widetext}
where the coefficient of $\left|ee\middle\rangle\middle\langle ee\right|$ is given by
\begin{widetext}
\begin{equation}
    \lambda=\frac{e^\frac{\pi\sqrt{1-4r^2}}{2r}\left(1+\sqrt{1-4r^2}\right)-e^{-\frac{\pi\sqrt{1-4r^2}}{2r}}\left(1-\sqrt{1-4r^2}\right)}{2e^\frac{\pi}{2r}\sqrt{1-4r^2}}\;.
\end{equation}
\end{widetext}
It shows that the imperfection is caused by the leakage of the doubly-excited state $\left|ee\right\rangle$. We can calculate the general distance by Equation~\ref{eq:normalization_distance}, and obtain
\begin{widetext}
\begin{equation}
    d_g\left(\mathcal{E}_Z,U_Z\right)=\frac{2e^{\frac{\pi}{2r}\left(1+\sqrt{1-4r^2}\right)}\sqrt{1-4r^2}-\left(1+e^{\frac{\pi}{r}\sqrt{1-4r^2}}\right)\left(1+\sqrt{1-4r^2}\right)+2}{2e^{\frac{\pi}{2r}\left(1+\sqrt{1-4r^2}\right)}\sqrt{1-4r^2}+\left(1+e^{\frac{\pi}{r}\sqrt{1-4r^2}}\right)\left(1+\sqrt{1-4r^2}\right)-2}\;.
    \label{eq:zeno_distance}
\end{equation}
\end{widetext}

\input{main.bbl}

\end{document}

%% file: main.bbl
%